\documentstyle[sprocl,epsf]{article}

\def\be{\begin{equation}}
\def\ee{\end{equation}}
\def\bea{\begin{eqnarray}}
\def\eea{\end{eqnarray}}
\begin{document}

\title{RECENT RESULTS IN THE CENTER VORTEX MODEL FOR THE INFRARED 
SECTOR OF YANG-MILLS THEORY\footnote{Talk presented by M.~Engelhardt.} }

\author{M. ENGELHARDT,\footnote{email: 
\verb+engelm@pion08.tphys.physik.uni-tuebingen.de+ } H. REINHARDT}

\address{Institut f\"ur theoretische Physik, Universit\"at T\"ubingen,\\
Auf der Morgenstelle 14, 72076 T\"ubingen, Germany}

\author{M. FABER}

\address{Institut f\"ur Kernphysik, Technische Universit\"at Wien,\\
A-1040 Vienna, Austria}

\maketitle\abstracts{A model for the infrared sector of $SU(2)$ Yang-Mills 
theory, based on magnetic vortices represented by (closed) random surfaces, 
is presented. The model quantitatively describes both confinement and the
topological aspects of Yang-Mills theory. Details (including an adequate
list of references) can be found in the e-prints hep-lat/9912003 and
hep-lat/0004013, both to appear in Nucl. Phys. B.}

Diverse nonperturbative effects characterize strong interaction physics.
Color charge is confined, chiral symmetry is spontaneously broken, and
the axial $U(1)$ part of the flavor symmetry exhibits an anomaly.
Various model explanations for these phenomena have been advanced;
to name but two widely accepted ones, the dual superconductor 
mechanism of confinement, and instanton models, which describe the 
$U_A (1)$ anomaly and spontaneous chiral symmetry breaking. However, no 
clear picture has emerged which comprehensively describes
infrared strong interaction physics within one common framework. The vortex 
model presented here~\cite{selprep,preptop} aims to bridge this gap. 
On the basis of a simple effective dynamics, it simultaneously reproduces 
the confinement properties of $SU(2)$ Yang-Mills theory (including 
the finite-temperature deconfinement transition), as well as the 
topological susceptibility, which encodes the $U_A (1)$ anomaly.
Remarks on the chiral condensate, an important point of investigation 
which has not yet been carried out, will be made in closing.

Center vortices are closed chromomagnetic flux lines in three-dimensional
space; thus, they are described by closed two-dimensional world-surfaces
in four-dimensional space-time. In the $SU(2)$ case, their magnetic flux is 
quantized such that they modify any Wilson loop by a phase factor $(-1)$
when they pierce an area spanned by the loop.
To arrive at a tractable vortex model, it is useful to compose the vortex 
world-surfaces out of plaquettes on a hypercubic lattice. The spacing of 
this lattice is a fixed physical quantity (related to a thickness of the 
vortex fluxes), and represents the ultraviolet cutoff inherent
in any infrared effective framework. The model vortex surfaces are
regarded as random surfaces, and an ensemble of them is generated using
Monte Carlo methods. The corresponding weight function penalizes curvature
by associating an action increment $c$ with every instance of two
plaquettes which are part of a vortex surface, but which do not lie 
in the same plane, sharing a link (note that several such pairs of 
plaquettes can occur for any given link). 

\begin{figure}
\centerline{
\epsfxsize=5.8cm
\epsffile{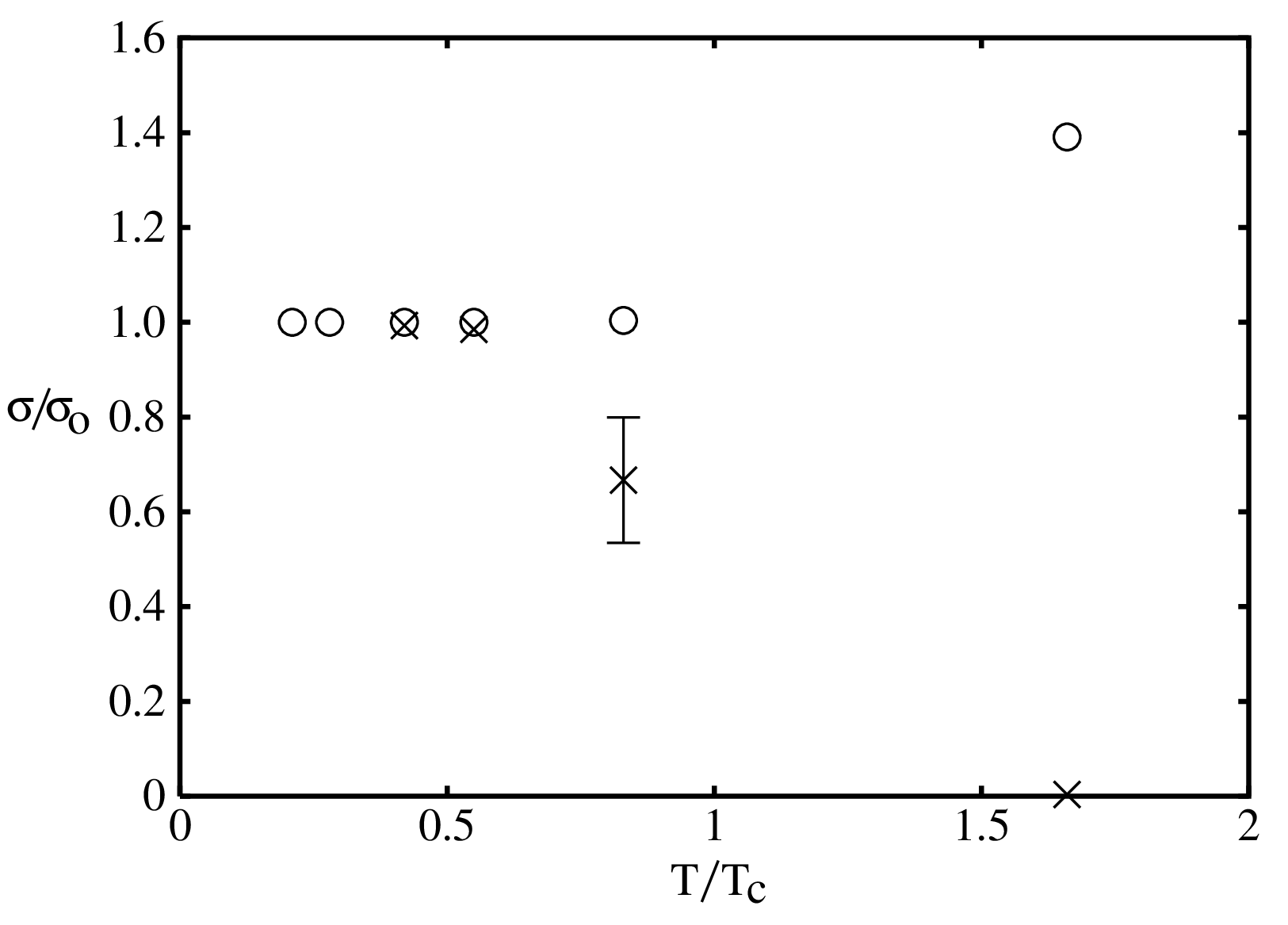} 
\hspace{0.1cm} 
\epsfxsize=5.8cm
\epsffile{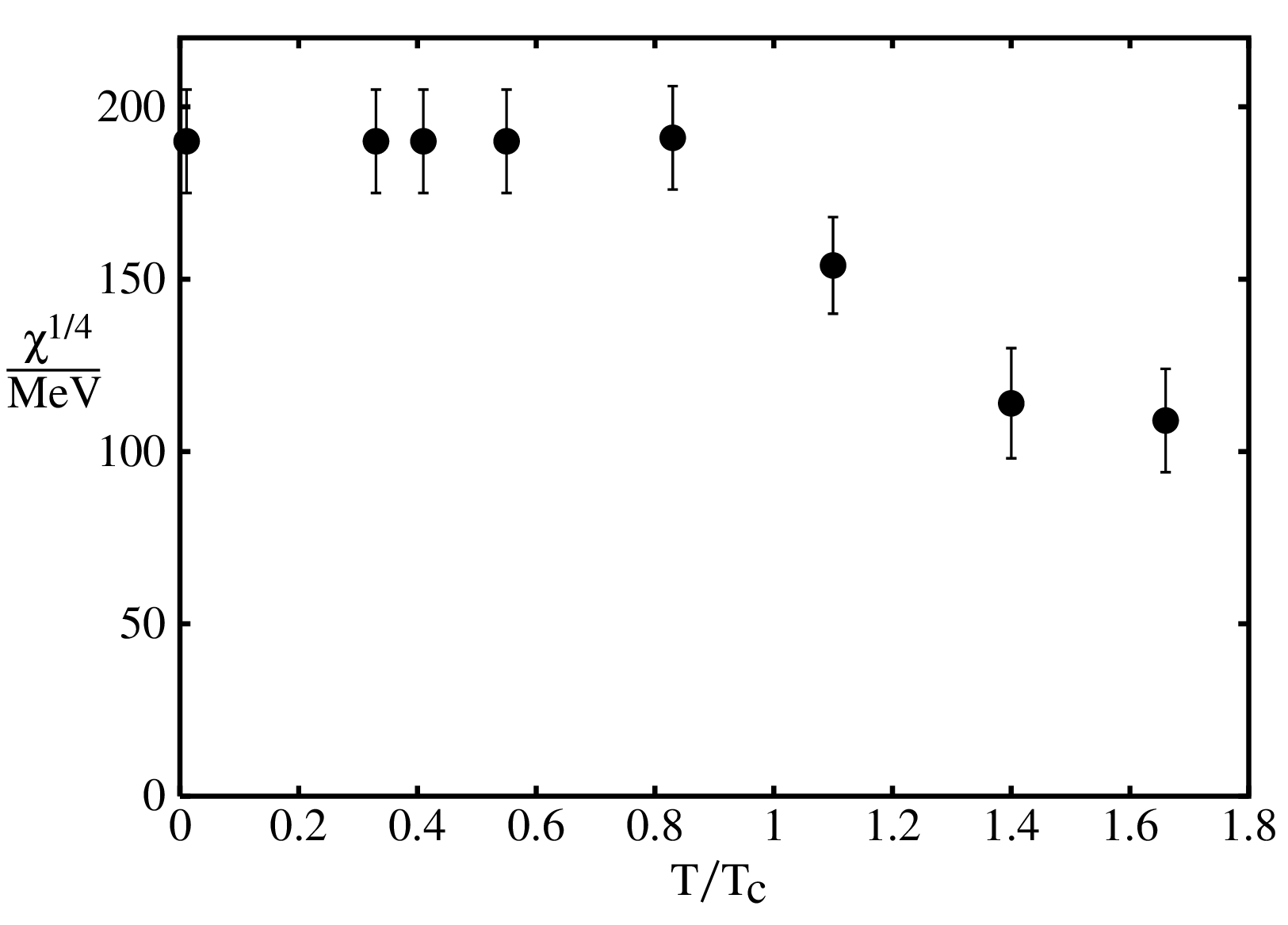} 
}
\caption{Observables in the random vortex surface model on
$16^3 \times N_t $ lattices, with $c=0.24$, as a function of temperature. 
Left: String tension between static color sources (crosses) and spatial 
string tension (circles). Whereas the quantitative behavior of the static 
quark string tension has largely been fitted using the freedom in the 
choice of $c$ (see text), the spatial string tension $\sigma_{s} $ is 
predicted. In the deconfined regime, it begins to rise with temperature; 
the value $\sigma_{s} (T=1.67 \, T_C ) = 1.39 \, \sigma_{0} $ corresponds to
within 1\% with the value measured in full $SU(2)$ Yang-Mills
theory.{\protect \cite{karsch} } Right: (Fourth root of) the
topological susceptibility; also this result is quantitatively compatible
with measurements in full Yang-Mills theory.{\protect \cite{digia} } }
\label{stt}
\end{figure}

Via the definition given above, Wilson loops (and, in complete analogy,
Polyakov loop correlators) can be evaluated in the vortex ensemble, and 
string tensions extracted. For sufficiently small curvature coefficient 
$c$, one finds a confined phase (non-zero string tension) at low 
temperatures, and a transition to a high-temperature deconfined phase. 
For $c=0.24$, the $SU(2)$ Yang-Mills relation between the deconfinement
temperature and the zero-temperature string tension,
$T_C /\sqrt{\sigma_{0} } =0.69$, is reproduced. When furthermore setting 
$\sigma_{0} =(440 \, \mbox{MeV} )^2 $ to fix the scale,
measurement of $\sigma_{0} a^2 $ yields the lattice spacing 
$a=0.39 \, \mbox{fm} $. The full temperature dependence of the
string tensions is displayed in Fig.~\ref{stt} (left).
Note that the confined and deconfined phases can alternatively be
characterized by certain percolation properties of the 
vortices.\cite{selprep,tlang}

Complementarily, also the topological properties of the Yang-Mills ensemble 
encode important nonperturbative effects. The topological charge $Q$ of a 
vortex surface configuration is carried by its singular 
points,\cite{cont,preptop} i.e.~points at which the set of tangent 
vectors to the surface configuration spans all four space-time directions 
(a simple example are surface self-intersection points). Since a vortex 
surface carries a field strength characterized by a nonvanishing tensor 
component associated with the two space-time directions locally orthogonal 
to the surface,\cite{cont} these singular points are precisely the points 
at which the topological charge density 
$\epsilon_{\mu \nu \lambda \tau } \, \mbox{Tr} \, F_{\mu \nu }
F_{\lambda \tau } $ is non-vanishing. In practice, implementing this
result for the hypercubic lattice surfaces used in the present model
involves resolving ambiguities~\cite{preptop} reminiscent of those
contained in lattice Yang-Mills link configurations. The resulting
topological susceptibility $\chi = \langle Q^2 \rangle /V$, where $V$
denotes the space-time volume under consideration, is exhibited in 
Fig.~\ref{stt} (right) as a function
of temperature. Taken together, the measurements in Fig.~\ref{stt}
show that the vortex model provides, within one common framework, a 
quantitative description not only of the confinement properties, but also
of the topological properties of the $SU(2)$ Yang-Mills ensemble. 

One obvious generalization of the present work is the treatment of
$SU(3)$ color. Also, the coupling of the vortex degrees
of freedom to quarks must be investigated, e.g.~whether the correct chiral
condensate is induced in the vortex background. In this respect, the 
vortex picture has an important advantage to offer. It is possible to
associate any arbitrary vortex surface with a continuum gauge 
field,\cite{cont} including the surfaces generated within the
present model. As a consequence, the Dirac operator, encoding quark
propagation, can be constructed directly in the continuum, and some
of the difficulties associated with lattice Dirac operators, such as
fermion species doubling, may be avoidable.

\section*{Acknowledgments} 
M.E.~and H.R.~acknowledge DFG financial support
under grants En 415/1-1 and Re 856/4-1, respectively. 
M.F.~is supported by Fonds zur F\"orderung der
Wissenschaftlichen Forschung under P11387-PHY.

\end{document}